\documentclass[fleqn, a4paper, 12pt]{article}
\usepackage{amsfonts, amssymb, amsmath, epsfig, graphicx, psfrag}

\begin{document}

\title{\bf A spatial model for social networks} 
\author{Ling Heng Wong\thanks{Corresponding author.  Current address: Department of Psychology, School of Behavioural Science, The University of Melbourne, VIC 3010, Australia.  e-mail: lingw@unimelb.edu.au.  Tel: +61 3 8344 6362.  Fax: +61 3 9347 6618.}, Phillipa Pattison, and Garry Robins\\
{\small\it Department of Psychology,}\\ {\small\it The University of Melbourne.}}
\maketitle
\bibliographystyle{plain}

\begin{abstract}
We study spatial embeddings of random graphs in which nodes are
randomly distributed in geographical space.  We let the edge
probability between any two nodes to be dependent on the spatial
distance between them and demonstrate that this model captures many
generic properties of social networks, including the ``small-world''
properties, skewed degree distribution, and most distinctively the
existence of community structures.
\end{abstract}



\section{Introduction}
Complex social networks arise in a wide range of contexts, for example
as corporate partnership networks \cite{Lazega01}, scientist
collaboration networks \cite{Newman02}, company director networks
\cite{RobinsAlexander2004}, film actors networks
\cite{AmaralScalcBarthelemy00}, sexual contact networks \cite{Morris97}, etc.
Indeed, a lot of attention has been given by both physical and social
scientists in recent years to model these networks so as to gain
better understandings of their general structures as well as their
various functions like information flow \cite{HubermanAdamic04},
locating individuals \cite{AdamicAdar04}, disease spread \cite{Morris97},
etc.  For a review of recent efforts, see for example
\cite{RobinsPattision05}, \cite{AlbertBarabasi02} and \cite{Newman03}.
While there is an apparent increase in the number of network models in
the literature, not all of these models have taken full advantage of
the sociological and psychological insights on how social networks may
be formed.

\subsection{Spatial characteristics of social ties}
The principle of \emph{homophily}, or in essence ``birds of a feather
flock together,'' has been firmly established by many empirical
studies \cite{McPhersonSmithLovinCook01}.  While we clearly tend to
befriend those who are like us, there are many situations where having
a lot of friends like us is simply because we are \emph{stuck} with
people who are like us in the first place.  For example if you are a
millionaire and all your friends are millionaires, it might simply be
because you were born into an elite family and live in an elite area
so you only know millionaires in your life, even though you do not
actively choose to befriend millionaires over non-millionaires.
Therefore, it is useful to divide homophily into two main types:
\emph{baseline} homophily and \emph{inbreeding} homophily
\cite{McPhersonSmithLovinCook01}.  Baseline homophily is attributed to
the fact that we have a
\emph{limited potential tie pool} due to factors like demography and
foci of activities \cite{Feld81}.  Inbreeding homophily is
conceptualised as any other kind of homophily measured over that
potential tie pool --- this may include homophily regarding gender,
religion, social class, education, and other intra-personal or
behavioural characteristics.  While many network models have taken
inbreeding homophily into account
\cite{WassermanGalaskiewicz84,RobinsElliottPattison01,NewmanGirvan03,Newman03a,RobinsJohnston04,vanDuijnSnijdersZijlstra04},
they have generally assumed that there are no baseline homophily
effects, i.e.\ the potential tie pool for all actors equals the
\emph{entire} population.  However, this is obviously not very
realistic and baseline homophily effects can potentially have profound
consequences on the structure of social networks.

A basic source of baseline homophily is the \emph{geographical space}.
As a matter of simple opportunity and/or the need to minimise efforts
to form and maintain a social tie \cite{Zipf49}, we can expect that we
tend to form ties with those who are geographically close to us.
Thus, intuitively, this creates a very strong constraint on our
potential tie pool.  In fact, there is ample empirical evidence that
demonstrates this claim.  The earliest studies of which we are aware
of date back to Festinger \emph{et al.} 
\cite{FestingerSchachterBack50} and Caplow and Forman
\cite{CaplowForman50} both on student housing communities.  The
results showed that in these rather homogeneous communities, spatial
arrangement of student rooms/units was an important factor in
predicting whether two dwellers have at least weak ties.  Many other
network studies also reached similar results, for example see
\cite{AthanasiouYoshioka73,BarrettCampbell99}.  More recently, Wellman
\cite{Wellman96} and Mok \emph{et al.}\
\cite{MokWellmanBasu04} re-analysed Wellman's earlier dataset on
Torontorian personal communities
\cite{WellmanCarringtonHall88,WellmanWortley90} and noted that
most personal friendships were indeed ``local,'' contrary to the
beliefs that recent technological advances have freed us from spatial
constraints.  For instance, in \cite{WellmanCarringtonHall88} it was
found that on average $42\%$ of ``frequent contact'' ties live within
a mere 1 mile radius of a typical person, while the rest of his/her
ties could be directed to anywhere in the rest of the world.

\subsection{General features of social networks}
Before embarking on specifying the model, we shall review some of the
general features of social networks.  Not many current models
simultaneously displays all of these.  We suggest that, by including
baseline spatial homophily into our network model, one can reproduce
all the following features, at least in broad terms:
\begin{enumerate}
 \item \textbf{Low tie density.} The number of possible ties in a
 network is theoretically quadratic to the number of actors, but most
 networks realise only a tiny fraction of these ties.  The cognitive
 ability of human places an upper bound on the number of ties one may
 maintain \cite{Dunbar92}.  On the other hand, other factors
 corresponding to baseline homophily can also play a role
 \cite{Feld81};

 \item \textbf{Short average geodesic distances.}  Geodesic distance
 between two actors is defined to be the length of the shortest
 connection between them.  In large social networks, it is believed
 that the typical geodesic distance between any two actors remains
 small.  This property was demonstrated empirically by Stanley Milgram
 in his classical experiment in the 1960s \cite{Milgram67},
 contributing to the popular saying that no one on this earth is
 separated from you by more than six ``handshakes'';

 \item \textbf{High level of clustering.} Clustering is defined to be
 the average probability that two friends of an actor are themselves
 friends.  Equivalently, it is a measure of how having a mutual friend
 will heighten the conditional probability that the two friends of an
 actor will be friends themselves.  In their well-known article
 \cite{WattsStrogatz98}, Watts and Strogatz demonstrated the
 importance of \emph{short-cuts} in social networks that
 simultaneously display high clustering and short average geodesic
 distances.  Such an idea of short-cuts dates back to Granovetter's
 arguments on the strength of weak ties \cite{Granovetter73};

 \item \textbf{Positively skewed actor degree distribution.} The
 degree of an actor is the number of social ties he/she has.  In many
 social networks, a majority of actors have relatively small degrees,
 while a small number of actors may have very large degrees.  This
 feature is displayed in a wide range of social networks.  While it is
 still debated whether generic social networks have power-law,
 exponential, or other degree distributions, or indeed whether there
 is any \emph{generic} distribution at all \cite{HandcockJones03},
 there is no doubt that degree distributions are in general positively
 skewed;

 \item \textbf{Existence of communities.}  In many cases, clustering
 does not occur evenly over the entire network.  We can often observed
 subgroups of actors who are highly connected within themselves but
 loosely connected to other subgroups which are themselves highly
 inter-connected.  We call these highly-connected subgroups
 \textit{communities} \cite{Newman04}.  A long tradition in social
 network analysis has developed a range of algorithms to identify
 these cohesive subsets of nodes \cite{WassermanFaust94}.
\end{enumerate}

\begin{figure}
  \begin{center}
	\psfrag{degree}{degree}
	\psfrag{frequency}{frequency}
	\psfrag{``deg-dist.txt''}{}
	\psfrag{1}{1}
	\psfrag{2}{2}
	\psfrag{3}{3}
	\psfrag{4}{4}
	\psfrag{5}{5}
	\psfrag{6}{6}
	\psfrag{7}{7}
	\psfrag{8}{8}
	\psfrag{9}{9}
	\psfrag{10}{10}
	\psfrag{11}{11}
	\psfrag{12}{12}
	\psfrag{13}{13}
	\psfrag{14}{14}
	\psfrag{15}{15}
	\psfrag{16}{16}
  \includegraphics[width=11cm]{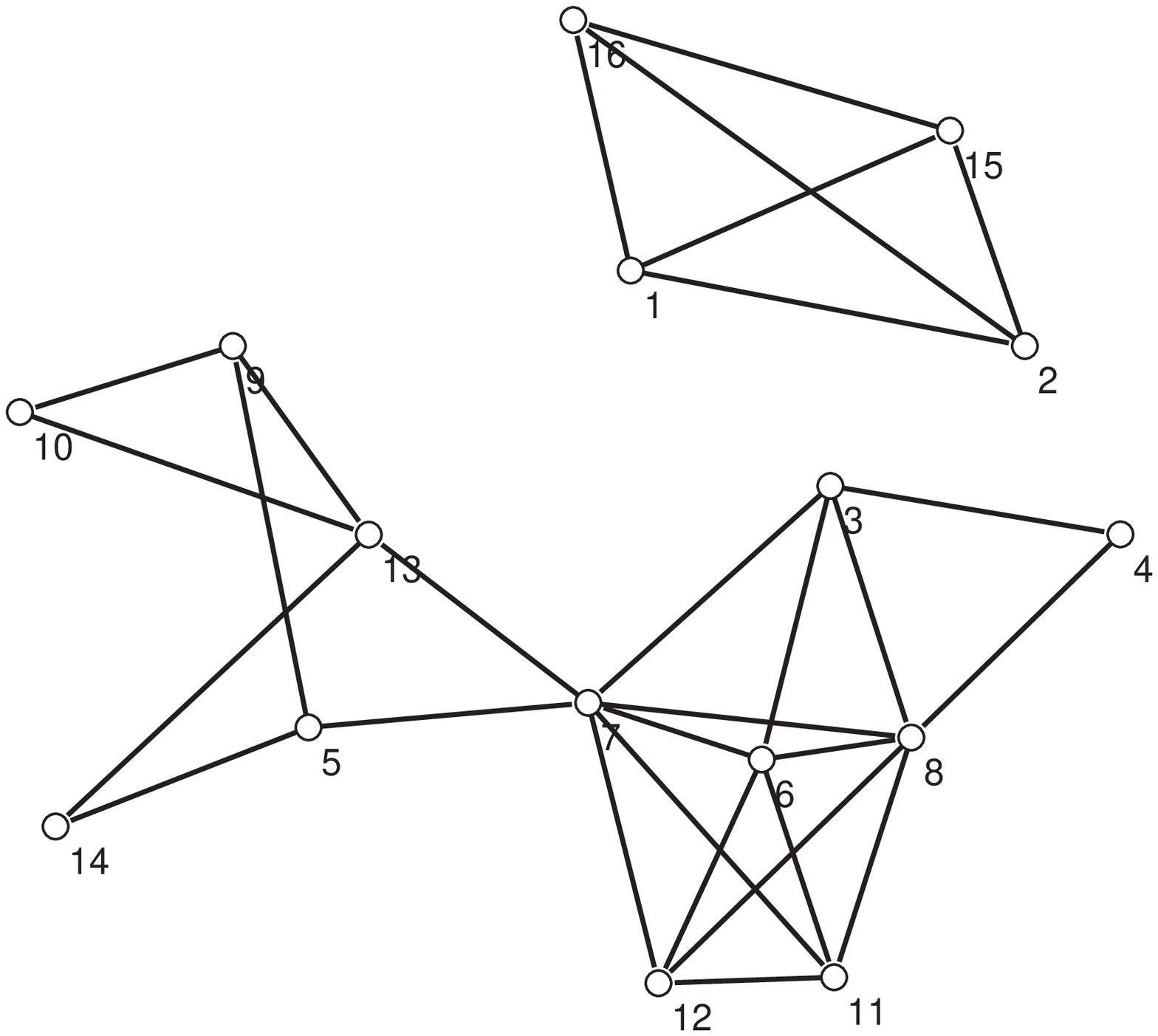} \\
  \end{center}
\caption{\label{fig:gamapos} The alliance network in Eastern Central Highlands of New Guinea.  Note that the numbers next to the nodes are for illustrative purpose only and the spatial arrangements of the nodes do not reflect the actual tribe locations.
}
\end{figure}

An example of a social network that displays all of the above
properties is a well-known alliance network of 16 tribes in the
Eastern Central Highlands of New Guinea \cite{Read54}\footnote{The
full data set is included as a sample data set in the standard social
network analysis program UCINET, which is available at {\tt
http://www.analytictech.com/ucinet.htm}.}.  The network is depicted in
Fig.\ \ref{fig:gamapos} where nodes correspond to tribes and ties
correspond to alliances between the relevant tribes.  First of all,
the density of the network is fairly low ($\rho=0.24$) given the small
size of the network.  The degree distribution is positively skewed
(skewness statistic = $0.99$).  The network is a ``small world'' in
which it has low median geodesic distance ($2$) and high level of
clustering coefficient (clustering coefficient = $0.63$).  Most
importantly, two distinct communities can be easily observed in Fig.\
\ref{fig:gamapos}: one is disjoint from the rest and is fully
connected (i.e.\ Nodes 1, 2, 15, and 16) and the other is highly
connected within itself (i.e.\ Nodes 3, 6, 7, 8, 11, and 12).

\subsection{Random graph models}
We here represent social networks by non-directed graphs.  A
non-directed graph is defined to be a pair $(\mathcal{V},
\mathcal{E})$, where $\mathcal{V}=\{v_1, v_2, \ldots, v_N\}$ is the
\emph{node set} denoting the individual actors in the network, and
$\mathcal{E}=\{e_1, e_2, \ldots, e_M\}$, with each edge $e_i$ being an
unordered pair of nodes $e_i = (v_r, r_s)$ ($r \neq s$ and $v_{r,s}
\in \mathcal{V}$), is the \emph{edge set} denoting the social ties
among the actors.  A compact way to represent a graph is through its
\emph{adjacency matrix} $\mathbf{X} = [x_{ij}]$, $i,j \in \{1, 2,
\ldots, N\}$ such that $x_{ij} = 1$ iff $(v_i, v_j) \in \mathcal{E}$,
otherwise $x_{ij} = 0$.  In general, the size of the set $\mathcal{V}$
(or equivalently the dimensions of $\mathbf{X}$) is fixed but whether
an edge $(v_i, v_j)^t \in \mathcal{E}$ (or equivalently $x_{ij} = 1$
in $\mathbf{X}$) is determined by a random process.  Such a random
process is defined so as to reflect the underlying social dynamics.

The simplest model for social networks is the Erd\"{o}s-R\'{e}nyi or
\textit{Bernoulli} random graph model, initiated independently by Paul
Erd\"{o}s and Alfred R\'{e}nyi \cite{ErdosRenyi59} and Anatol Rapaport
\cite{Rapoport53}, where the random process is a \emph{Bernoulli}
trial, i.e.
\begin{equation}
x_{ij} = \left\{ \begin{array}{ll} 1, & \mbox{ with }\mbox{probability
} p\\ 0, & \mbox{ with }\mbox{probability } 1-p\end{array} \right.,
\end{equation}
where the constant $p \in [0, 1]$ is called the \textit{edge
probability}.  In other words $x_{ij}$ are identically and
independently distributed (i.i.d.)\ Bernoulli random variables.  Due
to its simplicity, it is amenable to rigorous treatments.  It is
fairly straightforward to show that the average geodesic distance
between any two nodes is $\langle r_{ij} \rangle \sim \ln{N}$, a
feature that, as discussed above, resembles the property in some real
networks.  However the level of clustering in this model can be shown
to vanish as $N \to \infty$ and it is one of the major short-comings
of this simple model in modelling social networks.  Further, Erd\"{o}s
and R\'{e}nyi showed that there is a critical edge probability $p_c
\approx N^{-1}$ at which there always exists a connected component
containing a significant proportion of nodes in the network (almost
surely) \cite{Bollobas85}.  Such component is known as the giant
component.  For an extensive review of these results refer to
Bollob\'{a}s \cite{Bollobas85} and Janson
\textit{et al.} \cite{JansonLuczakRucinski99}.

\subsection{The overview}
The main advantage of the Erd\"{o}s-R\'{e}nyi model is its simplicity.
Although it does not predict some of the generic features outlined
above, it serves as a good foundation to build more realistic models.
In this paper, we study a generalisation of Erd\"{o}s-R\'{e}nyi random
graph model for social networks which incorporates a simple
\emph{baseline spatial homophily} effect in the formation of
individual network ties.  We note that this class of models is for a
\emph{single snapshot} of a network, thus \emph{temporal network
dynamics} are not taken into account.  In Section 2, we shall specify
the model and examine some of its basic properties.  In Section 3, we
shall outline our simulation methods and discuss the main results.  In
Section 4, we shall demonstrate the application of our model in a
particular social network.  In Section 5, we shall discuss the
implications of the results and describe ongoing research on this and
more generalised models.

\section{Spatial random graph model}
First of all, we embedded the nodes of graphs in the Euclidean space
$\mathbb{R}^2$ with a \emph{distance} function $d$ defined to map any
unordered pair of nodes to a real number, i.e. $d: \mathbb{R}^2
\times \mathbb{R}^2 \to \mathbb{R}$.  Further, $d$ satisfies the standard
triangle inequalities and the positivity condition.  Every node $v_i$
is assigned a coordinate $(x_i, y_i)^t$ according to some spatial
distribution and we define $\chi = \{(x_1, y_1)^t, (x_2, y_2)^t,
\ldots \}$ to be the location vector that specifics the spatial
locations of \emph{all} nodes in a network.  Further we shall use the
shorthand $d_{ij} = d((x_i, y_i)^t, (x_j, y_j)^t)$ to denote the
distance between nodes $v_i$ and $v_j$.  Next, we assume that the
spatial locations of the nodes are randomly scattered in space (and
therefore all locations are mutually independent as well).  In other
words, we assumed that the points are distributed in space according
to a homogeneous Poisson point process (see for example
\cite{CoxIsham80,Kingman93}).  Recall that a Poisson point process with
rate $\rho < \infty$ in the $d$-dimensional Euclidean space
$\mathbb{R}^d$ is a process such that:
\begin{itemize}
  \item for all disjoint subsets $A_1, A_2,\ldots,A_k \subset
  \mathbb{R}^d$, the random variable denoting the number of point in
  the each subset, $N(A_1), N(A_2), \ldots, N(A_k)$, are independently
  distributed; 
  \item $N(A)$ has Poisson distribution; and 
  \item $E[N(A)] = \rho |A|$ for all $A \subset \mathbb{R}^d$.
\end{itemize}

To model the baseline homophily effect, we let the edge probability
between two nodes to be dependent on the spatial distance between
them.  That is, given a $\chi$, $P(x_{ij} = 1 | \chi) = f(d_{ij})$,
where $f:\mathbb{R} \to [0,1]$.  Motivated by the discussion in the
last section, we shall consider the case where $f$ is a simple step
function, i.e.
\begin{equation}
P(x_{ij} = 1 | \chi) = \left\{ \begin{array}{ll} p+p_b, & \mbox{if }
d_{ij} \le H, \\ p-\Delta, & \mbox{if } d_{ij} > H, \end{array}
\right.  \label{eq:modeldef}
\end{equation}
where $p$ is the average density of the network, $H$ is the
\emph{neighbourhood radius}, $p_b$ is the \emph{proximity bias} which
specifies the sensitivity to geographical space by the actors on
establishing social links.  Thus $p_b$ probabilistically controls the
locations of potential tie pool.  Of course we have assumed that such
a spatial sensitivity is the same for all actors in the network.
Further, $\Delta = \Delta(p_b, H | \chi)$ is some correction term.
The correction term $\Delta$ is introduced to maintain the expected
average density to be a constant $p$, given $\chi$, for all feasible
values of $H$ and $p_b$, i.e.
\begin{equation}
E\left[\left.{\frac{1}{N-1}\sum_{i<j}x_{ij}}\right|\chi\right] = p,
\end{equation}
and there is no other substantive purpose.  Without maintaining
constant density, it will be difficult to isolate the effects of $p_b$
and $H$ on the graph's structural properties because of the
confounding effects from the varying expected number of edges in the
model.

\begin{figure}
 \begin{center}
 \includegraphics[width=13cm,angle=-0]{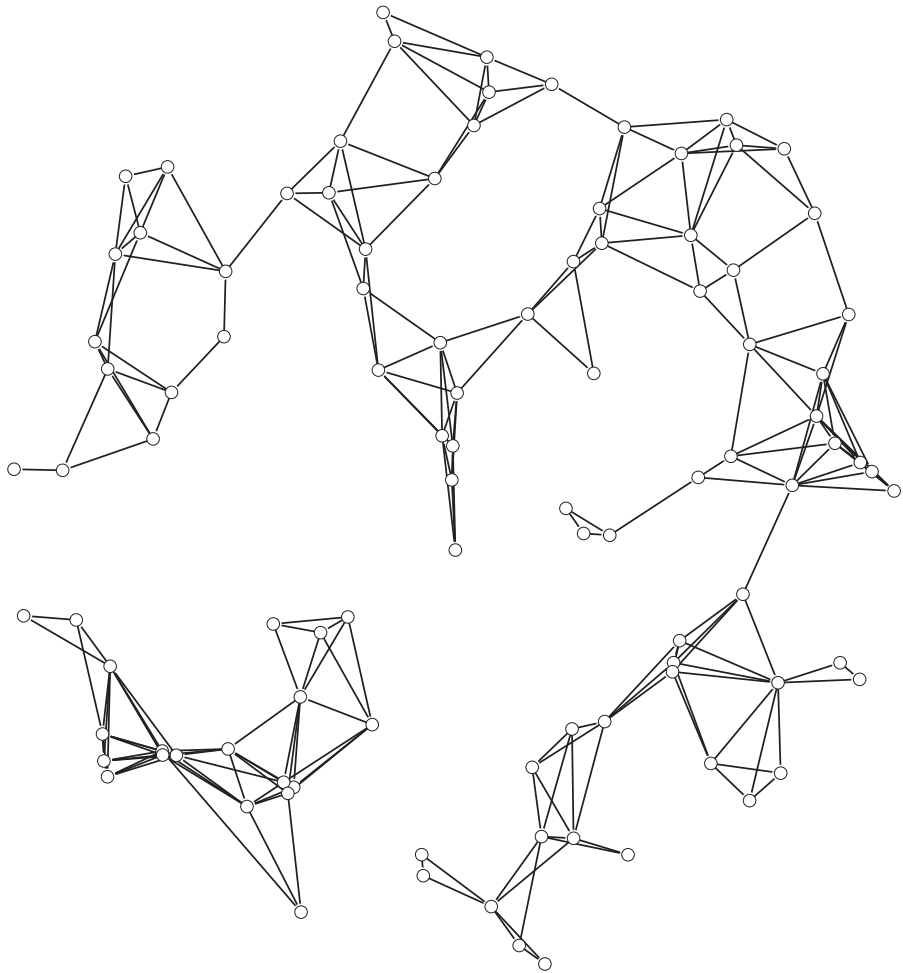}
 \end{center}
\caption{\label{fig:srg-example}A 100-node graph ($p=0.05$, $p_b=0.95$, and $H=3/2$) taken from the Markov Chain Monte Carlo simulation in Section 3.  Points are distributed according to a Poisson point process.  
}
\end{figure}

To calculate $\Delta$, we first determine the number of all possible
edges shorter than the neighbourhood radius $H$ in the network
embedded in $\chi$, call this number $S_{\le H}(\chi)$; when $N$ is
sufficiently large (ignoring boundary effects), $S_{\le H}(\chi)
\approx N\pi\rho^2/2$.  The number of possible edges longer than the
neighbourhood radius is therefore $S_{> H}(\chi) = \binom{N}{2} -
S_{\le H}(\chi)$.  By definition the expected density within
neighbourhoods is $p + p_b$, so the expected number of all realised
edges within all neighbourhoods is $(p+p_b)S_{\le H}$.  It follows
that, in order to maintain density to equal $p$, the expected number
of realised edges outside the neighbourhood ought to be $\binom{N}{2}p
- (p+p_b)S_{\le H}$.  As a result,
\begin{equation}
p - \Delta = \frac{1}{S_{> H}}\left[\binom{N}{2}p - (p+p_b)S_{\le
H}\right] = p - \frac{S_{\le H}}{\binom{N}{2} - S_{\le
H}}p_b. \label{eq:delta}
\end{equation}
\footnote{If the expected number of nodes within the neighbourhood
is the same across all nodes (e.g.\ when we have a homogeneous point
process with periodic boundary condition), then let the expected
number of nodes within a node's neighbourhood (not including itself)
be $E[s_{\le H}]$, we can then repeat the above and arrive at a
simpler form for the correction term,
\begin{equation}
\Delta = \frac{E[s_{\le H}]}{(N-1)-E[s_{\le H}]}p_b.
\end{equation}
}  Since the probability $P(x_{ij} = 1|\chi)$ is bounded by 0 and 1, $0
\le p+p_b \le 1$ and $0 \le p - \Delta \le 1$.  These conditions
defines upper and lower bounds for $p_b$ given a $p$ and $H$.  We note
that at the a particular set of values where $p = \left[\binom{N}{2} -
S_\le(\chi)\right] / \binom{N}{2}$ and $p_b = 1- p$, i.e. when there
is probability 1 that all pair of nodes less than distance $H$ apart
will be joint and probability 0 otherwise, this model becomes the
so-called \emph{random geometric graph model}
\cite{Penrose03}\footnote{We also later become aware of a
generalisation of the random geometric graph model called the
\textit{random connection model} where the edge probability is taken
to be a general decreasing function of spatial distance, $g$
\cite{Penrose91}.  In \cite{Penrose91}, some rigorous results regarding the giant cluster on the model has been obtained, however, the main variable of their model is the Poisson rate $\rho$ in space while keeping $g$ general.
However here our focus on the function $g$.}.  A typical instance of
spatial random graph can be found in Fig.\ \ref{fig:srg-example}.

It is convenient to express our model in the exponential form which
fits into the exponential random graph network modelling framework.
More details of the framework can be found in Wasserman and Pattison
\cite{FrankStrauss86,WassermanPattison96}.  The modelling framework has recently received renewed attention in the physics community, see
\cite{BurdaJurkiewiczKrzywicki04} and \cite{ParkNewman04} .  Let $x$
be an instance of random graph with spatial locations of nodes
specified by $\chi$.  Then the general form of the probability
function for obtaining $x$ is given by
\begin{equation}
P(X=x | \chi) = \frac{1}{Z(\chi)} \exp{\left[-\mathcal{H}(x | \chi)\right]}, \label{eq:expform}
\end{equation}
where $\mathcal{H}(x | \chi)$ is the Hamiltonian of graph $x$ given the
locations of nodes specified by $\chi$ and
\begin{equation}
Z(\chi) = \sum_x \exp{\left[-\mathcal{H}(x|\chi)\right]}
\end{equation} 
is the partition function of the model.  The Hamiltonian can take any
(sensible) form to reflect the dependency of edges.  The simplest
choice is $-\mathcal{H}(x) = \theta_0L(x|\chi)$, where $L(x|\chi) =
L(x)$ is the number of edges in the graph $x$ independent of $\chi$
and $\theta_0$ is its associated parameter.  By tuning $\theta_0$, we
can change the expected density of a typical graph in the model.  This
gives us the classical Erd\"{o}s-R\'{e}nyi random graph model
\cite{ErdosRenyi59} as introduced in Section 1.  To define
$\mathcal{H}(x|\chi)$ for our simple spatial model, we first define
$L_\le(x | \chi) = \sum_{i<j, d_{ij}\le H}x_{ij}$ and $L_>(x |
\chi) =\sum_{i<j, d_{ij} > H}x_{ij}$ be the number of edges shorter
and longer than $H$ respectively, then the Hamiltonian for our model
can be written as
\begin{equation}
-\mathcal{H}(x | \chi) = \theta_\le L_\le(x | \chi) + \theta_> L_>(x | \chi),
\end{equation}
where $\theta_\le$ and $\theta_>$ are parameters whose values can be
calculated directly from Eq.\ \ref{eq:modeldef}, i.e.
\begin{equation}
\theta_\le = \mbox{logit}(p+p_b), \qquad \mbox{ and } \qquad \theta_> = \mbox{logit}(p-\Delta),
\end{equation}
where $\mbox{logit }q = \log[q/(1-q)]$.  Equivalently, we can also
write the Hamiltonian in another form: $-\mathcal{H}(x | \chi) =
\theta L(x|\chi) + \theta' L_{\le}(x | \chi)$, where $L(x|\chi) =
\sum_{i < j}x_{ij}$ $(= L_\le(x | \chi) + L_>(x | \chi))$, $\theta =
\theta_>$ and $\theta' = \theta_\le - \theta_>$.  In the following however, we
shall only use the former form of the Hamiltonian.

The simplicity of the Hamiltonian allows us to write down the
partition function in closed form,
\begin{eqnarray*}
Z_N(\chi) = \sum_x \exp{\left[-\mathcal{H}(x)\right]} &=& \sum_x
\exp\left(\theta_\le L_\le(x) + \theta_> L_>(x)\right) \\ &=& \sum_x
\exp\left(\theta_\le
\sum_{i<j,d_{ij} \le H} x_{ij} + \theta_> \sum_{i<j,d_{ij} > H} x_{ij}
\right) \\  &=& \left(1 + e^{\theta_\le}\right)^{S_\le(\chi)} \left(1 +
e^{\theta_>} \right)^{\binom{n}{2} - S_\le(\chi)}.
\end{eqnarray*}
Using this explicit form, one can double-check the constant density in
our model for all $p_b$ and $H$ given $\chi$:
\begin{eqnarray*}
  \binom{N}{2}^{-1}\langle L(x | \chi) \rangle &=& \binom{N}{2}^{-1}\frac{1}{Z(\chi)}
  \sum_x \left[L_\le(x|\chi) + L_>(x|\chi)\right]\exp\left[-\mathcal{H}(x |
  \chi)\right] \\ &=& \binom{N}{2}^{-1}\frac{1}{Z(\chi)}\left(\frac{\partial Z(\chi)}{\partial
  \theta_\le} + \frac{\partial Z(\chi)}{\partial\theta_>}\right) \\
  &=& \binom{N}{2}^{-1} \left[S_\le(\chi)(p+p_b) + S_>(\chi)(p-\Delta)\right] = p,
\end{eqnarray*}
which is what we expected.  In theory, we can calculate any
statistical average statistical quantities $\langle Q(x|\chi)
\rangle$, for example the average clustering coefficient etc, from
$Z(\chi)$ by adding an auxiliary term in the Hamiltonian $\Delta
\mathcal{H}(\chi) = yQ(x|\chi)$.  Then,
\begin{eqnarray}
\langle Q(x|\chi) \rangle &=& \frac{1}{Z(\chi)}\sum_x Q \exp{\left[-\mathcal{H}(x|\chi) - yQ(x|\chi)\right]} \\ &=& \frac{1}{Z(\chi)}\left.\frac{\partial Z(\chi)}{\partial y}\right|_{y = 0} \label{eq:intheory}
\end{eqnarray}
However, as is the case in many other statistical mechanics models,
equations like Eq.\ \ref{eq:intheory} are very difficult to evaluate
exactly as a general approach in not yet available.  In this study, we
resort to using numerical simulations to explore the properties our
model.

\section{Simulation results and discussions}
The Poisson rate $\rho$ and the neighbourhood radius are relative to
each other, so we shall always fix $\rho = 1$ and vary $H$ only.  On
the other hand, the choice of the value of $H$ is not crucial as long
as $H$ is sufficiently large.  When $H$ is too small, the vast
majority of ties connected to a node are inevitably from the outside
of the neighbourhood, therefore, the model behaves as the simple
Erd\"{o}s-R\'{e}nyi random graph model.  Here we fix $H=3/2$.  The main
program of simulations below is to vary the proximity bias $p_b$ and
investigate the effects on the overall structures of the graphs.  We
used a Markov Chain Monte Carlo method outlined in Snijders
\cite{Snijders02} for all our simulations.  Each individual data point
below is a result of a simulation run ($250,000$ Markov iterations) of
random graphs with 100 nodes on a fixed $\chi$.  The burn-in phase for
each run is about $30,000$ iterations and statistics from this phase
were removed before further analysis.  The estimates of the statistics
are then calculated as the simple averages in the post-burn-in phase.
And we collect the estimates for six realisations of $\chi$.

\subsection{Number of short and long edges}
\begin{figure}
 \begin{center} \psfrag{LONGE}{\tiny{number of long edges}}
 \psfrag{SHORTE}{\tiny{number of short edges}}
 \includegraphics[width=11cm,angle=-0]{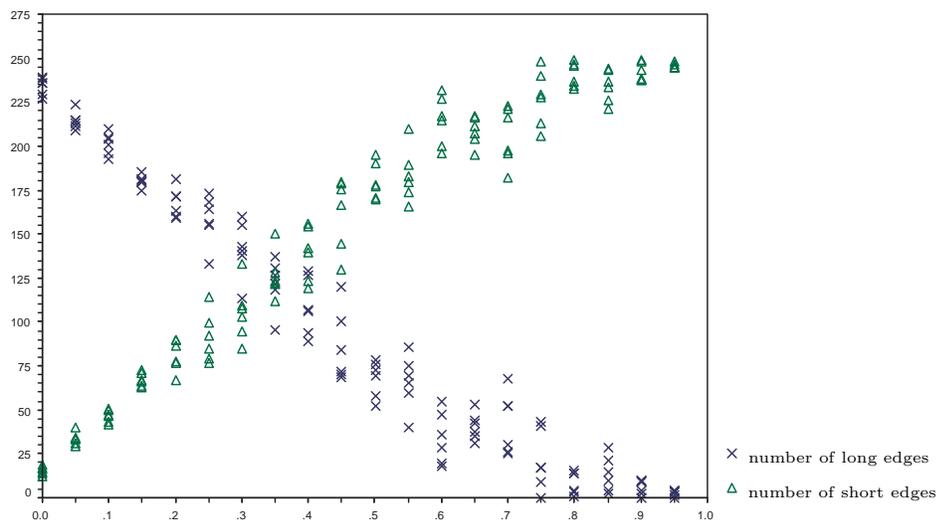} \end{center}
\caption{\label{fig:H15longshort} A plot of average number of short edges $\langle L_\le(x) \rangle$ and average number of long edges $\langle L_>(x) \rangle$ versus $p_b$.  }
\end{figure}

In Fig.\ \ref{fig:H15longshort}, we plot the average number of edges
shorter than ($\langle L_\le(x) \rangle$) and longer than ($\langle
L_>(x) \rangle$) the neighbourhood radius; we called them \emph{short}
and \emph{long} edges respectively.  When $p_b=0$, the opportunity tie
pool of each node equals the entire population (except itself).  In
this case, since there are many more potential long edges than short
edges in the graph, the graphs are on average dominated by long edges.
As $p_b$ increase, the potential tie pool of each node concentrate
more and more on the population geographically close, so short edges
dominate.  The two scatter plots in Fig.\ \ref{fig:H15longshort}
clearly display linear opposing trends.  The linearity is simply a
result of our model definition, refer to Eq.\ \ref{eq:modeldef}.  Also
as $p_b$ increases, there is growing variance of statistics for fixed
$p_b$.  It is because as the difference between short edge probability
and long edge probability grows, the graphs are more and more
dependent on the configuration of the \emph{specific instance} of the
point process. However, as $p_b$ approaches $1-p$, the variability
decreases again.  It is because the large $p_b$ values place
significant constraints on the feasible instances of the point process
and thus only a small number of instances $\chi$ are feasible for
large $p_b$ (see Eq.\ \ref{eq:delta}).

\subsection{Small-world properties}
\begin{figure}
  \begin{center} 
	\psfrag{PL}{\tiny $\langle L(X) \rangle$}
 	\psfrag{C}{\tiny $\langle C(X) \rangle$}
	\psfrag{PB}{\tiny $p_b$}
\includegraphics[width=11cm]{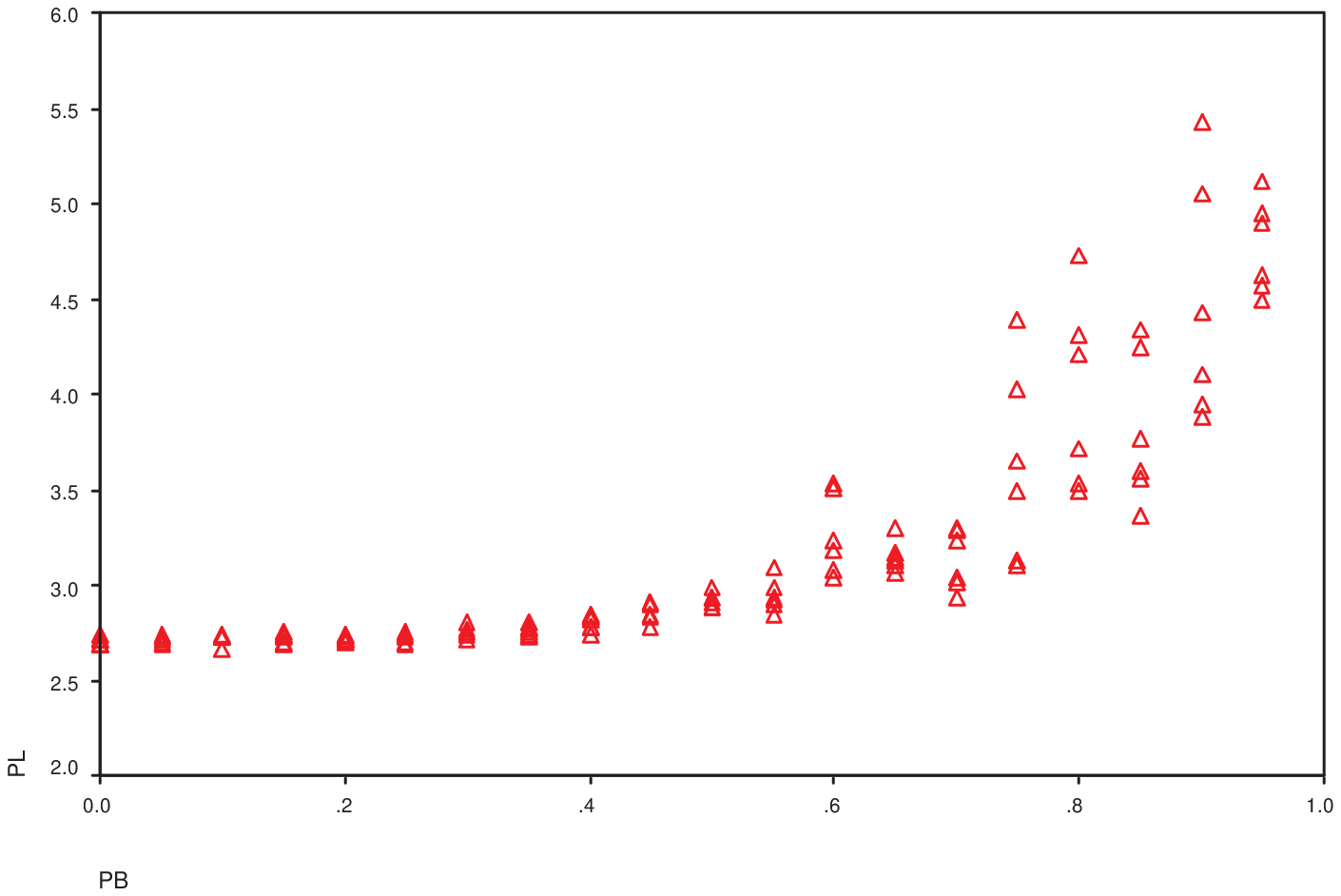}
  \includegraphics[width=11cm]{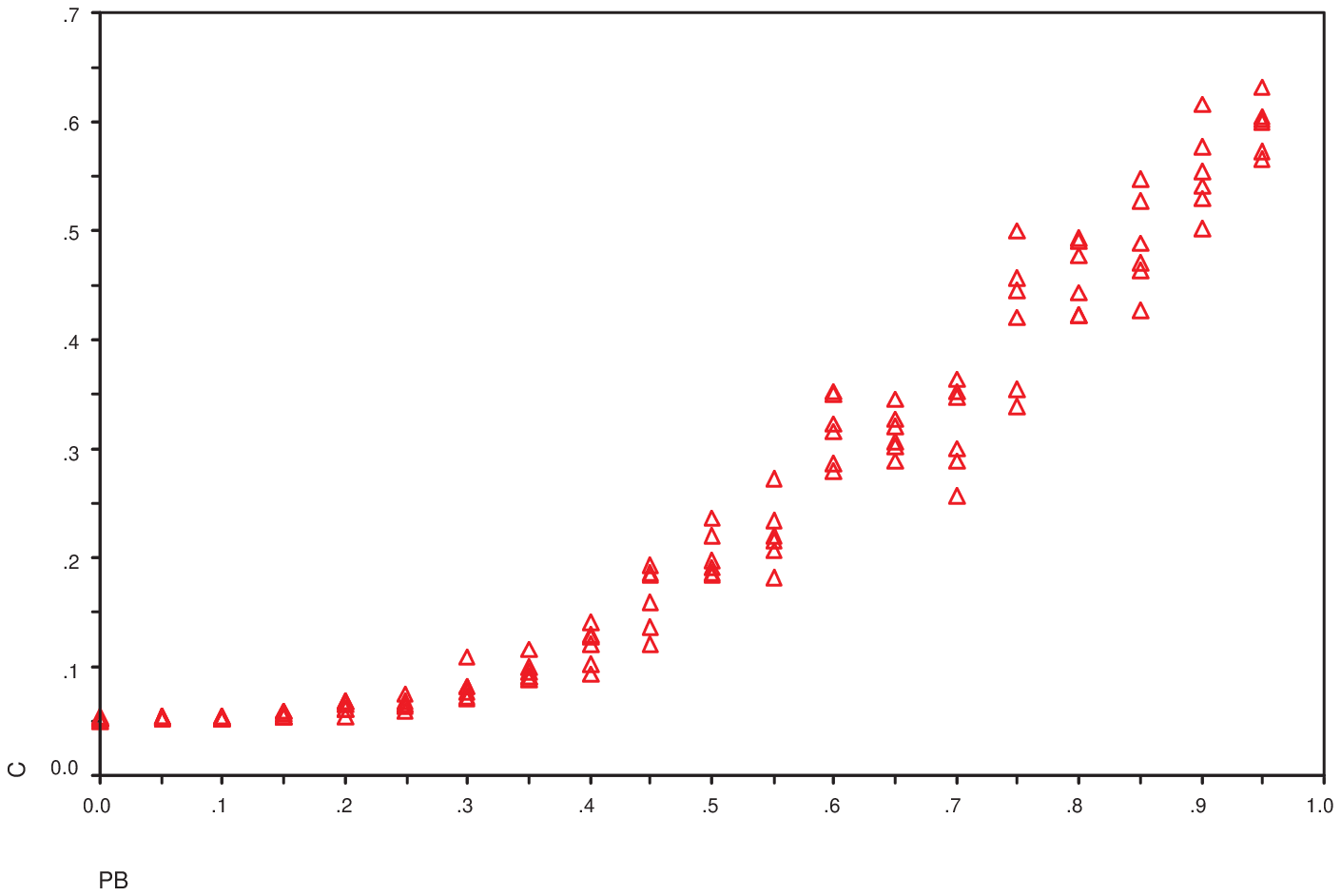} \end{center}
\caption{\label{fig:H15plc} A plot of average geodesic distance $\langle L(X) \rangle$ (upper graph) and average clustering coefficient $\langle C(X) \rangle $ (lower graph) against $p_b$.
}
\end{figure}

Let us now investigate the effect of $p_b$ on the global structure of
the graphs.  First let us define the \emph{geodesic distance}, or
graph distance, between two nodes.  Note that this distance is
independent of the spatial distance between the nodes involved.  Given
a graph $X=\{x_{ij}\}$, let the geodesic distance between $v_i$ and
$v_j$, $l_{ij}(X)$, be length of the shortest \emph{self-avoiding
path} connecting $v_i$ and $v_j$.  If $v_i$ and $v_j$ are
disconnected, we assign $l_{ij} = \infty$.  Now, given a $\chi$,
define
\begin{equation}
  w(X) = \left[\binom{N}{2}^{-1}\sum_{i<j} l_{ij}^{-1}\right]^{-1},
\end{equation}
to be the (harmonic) mean of $l_{ij}(X)$ over all possible pairs on
nodes in $X$.  $w(X)$ is a simple measure of \emph{connectivity} of
$G$.  Refer to the upper part of Fig.\ \ref{fig:H15plc} for a plot of
$\langle w(X) \rangle$ against $p_b$.  From the plot, $\langle w(X)
\rangle$ remains small for a large range of $p_b$.  Then there appears to
be a critical $p_b$, as in the Watts-Strogatz model
\cite{WattsStrogatz98}, that $w$ dramatically increases.  This
indicates a switch from the \emph{simple random graph} regime, where
``short-cuts'' between neighbourhood are abundant, to the \emph{random
geometric graph} regime, where most edges are within each actor's
neighbourhood.

Further, we study the level of clustering in the model.  Let $t_1(X)$
be the number of independent $3$-cycles, or triangles, in graph $X$,
i.e. $t_1(X) = \sum_{i<j<k}x_{ij}x_{jk}x_{ik}$.  Also let $s_2(X)$ be
the number of $2$-stars in graph $X$, i.e. $s_2(X) =
\sum_{i<j<k}x_{ij}x_{ik}$.  Then we define the global clustering
coefficient to be
\begin{equation} 
   C(X) = \frac{3t_1(X)}{s_2(X)}.  
\end{equation} 
$C(X)$ measures the overall level of clustering in a network or in
other words how much on average an actor's friend's friends are also
the actor's friends.  By definition, $0 \le C(X) \le 1$ for all $X$.
Refer to Fig.\ \ref{fig:H15plc} for a plot of the average $\langle C
\rangle$ over all $X$ in the ensemble as we changes $p_b$.  This model
also display similar behaviours as in the Watts-Strogatz model, and
there is a steady increase in its value as $p_b$ increases.  The
source of this clustering is entirely \emph{spatial}, i.e.\ triangles
are likely to be formed simply because of the fact that the involved
nodes are closed to each other spatially.

Overall, there is a range of $p_b$ where the model display
\emph{simultaneously} relatively low average path length \emph{and}
significant level of clustering --- a signature of a ``small-world''
model.

\subsection{Community structures}
\begin{figure}
  \begin{center}
	\psfrag{KTR}{\tiny $\langle T_{\lambda}(X) \rangle$}
	\psfrag{PB}{\tiny $p_b$}
    \includegraphics[width=11cm]{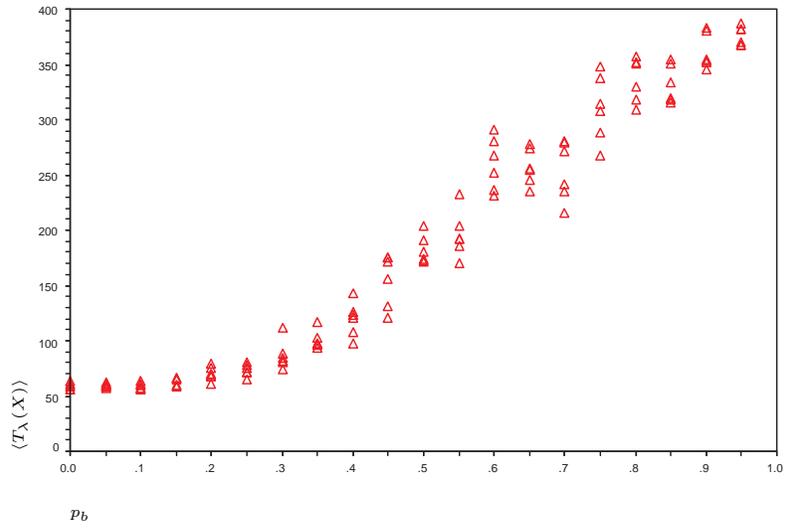} 
  \end{center}
\caption{\label{fig:H15ktri-m-tri} A plot of average $k$-triangle statistics $\langle T_{\lambda}(X) \rangle$ \cite{SnijdersPattisonRobinsHandcock05} for $\lambda=2$ against $p_b$. }
\end{figure}

Although the clustering coefficient defined above measures the
\emph{overall} level of clustering in a graph, there is the question
of whether clustering is distributed evenly over the entire graph on
average, in other words, do the triangles in the graph tend to clump
together or not?  The ``clumpiness'' of triangles can be measured by
the number of \emph{higher-order triangles}.  A $2$-triangle is
defined to be the combination of two triangles sharing a common edge
(which is called the base edge).  In general, a $k$-triangle is
defined to be the combination of $k$ triangles all sharing a common
base edge and let $t_k(X)$ its number in $X$.  Let $g_{ij}(G)$ be the
number of two-paths connecting $v_i$ and $v_j$, then a useful and
convenient way to combine all $t_k(G)$ into a single measure is as
follows:
\begin{eqnarray}
T_\lambda(G) &=& 3t_1(G) - \frac{t_2(G)}{\lambda} +
\frac{t_3(G)}{\lambda^2} - \ldots +
(-1)^{n-3}\frac{t_{n-2}(G)}{\lambda^{n-3}} \\ &=& \lambda \sum_{i < j}
x_{ij} \left[1 - \left(1-\frac{1}{\lambda}\right)^{g_{ij}(G)}\right],
\end{eqnarray}
where $\lambda$ is an arbitrary constant.  For a detailed discussion
of the motivation for this definition, see
\cite{SnijdersPattisonRobinsHandcock05}.  A plot of $\langle
T_{\lambda}(X) \rangle$ for $\lambda=2$ against $p_b$ can be found in
Figure \ref{fig:H15ktri-m-tri}.  There we can see a steady increase in
$T_{2}(X)$ as $p_b$ increases.  This suggests that graphs have higher
tendencies on average to form clumps of triangles --- we call these
clumps \emph{communities} --- for large $p_b$.  Inspection of
instances of graphs when $p_b$ is large suggests that the delineation
of communities are determined by large gaps in the spatial
distribution of the nodes.  This happens in spite of that fact that
the nodes are distributed uniformly randomly in space.  This phenomena
can be compared with the observation that communities are likely to be
separated by large streets, railroad tracks, etc \cite{Logan78}.

\begin{figure}
  \begin{center}
	\psfrag{GIANT}{\tiny $\langle \Gamma(X) \rangle$}
 	\psfrag{PB}{\tiny $p_b$}
	\includegraphics[width=11cm]{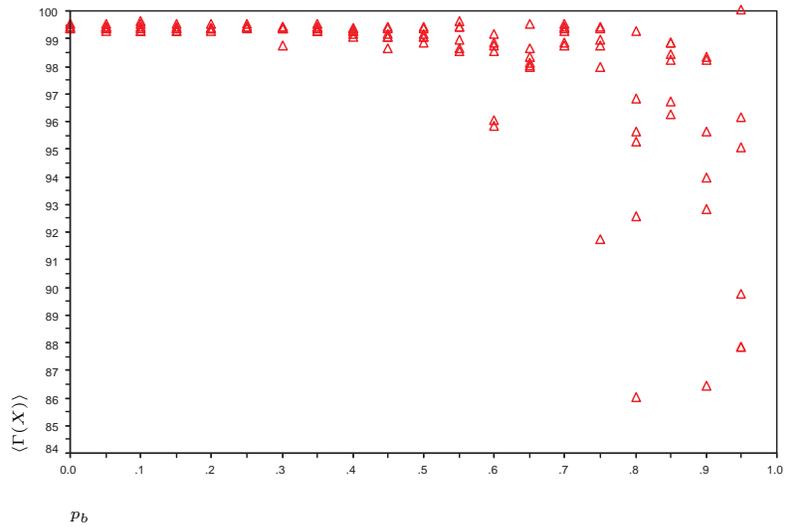} \end{center}
\caption{\label{fig:H15gc} A plot of average largest component size $\langle \Gamma(X) \rangle$ against $p_b$.
}
\end{figure}

Now that we have shown that a typical graph in the model is likely to
have strong community structures when $p_b$ is reasonably large.  One
possible implication of having such kind of structures is that the
whole graph can be composed of \emph{disjoint} communities.  To detect
whether it is the case, we consider the average size of the largest
connected component over the ensemble.  We define the \emph{component
size} containing node $v_i$, $\gamma_{i}(X)$, to be the number of
nodes with finite path length from $v_i$ (including itself),
i.e. $\gamma_{i}(G) = \sum_{i < j} \delta(l_{ij} <
\infty)$, where $\delta$ is the normal delta function.  Let
\begin{equation}
\Gamma(X) = \max_{v_i}\gamma_{v_i}(X)
\end{equation}
to be the number of nodes of $X$ in \emph{largest} component over all
$v_i$.  We have plotted the $\langle \Gamma(X) \rangle$ against $p_b$
in Fig.\ \ref{fig:H15gc}.  When $p_b$ is small, a vast majority of
nodes is contained in the largest component.  As $p_b$ increases,
there is a higher chance that the largest component no longer contains
most nodes, leaving a significant proportion of nodes in smaller
isolated components.  The instance of network in Fig.\
\ref{fig:srg-example} is an example of such a situation.

\subsection{Actor degree distributions}
\begin{figure}
  \begin{center}
    \psfrag{prob}{\tiny prob}
    \psfrag{PDEG08}{\tiny $p_b = 0.8$}
    \psfrag{PDEG04}{\tiny $p_b = 0.4$}
    \psfrag{PDEG00}{\tiny $p_b = 0.0$}
    \psfrag{PB}{\tiny $p_b$}
    \includegraphics[width=11cm]{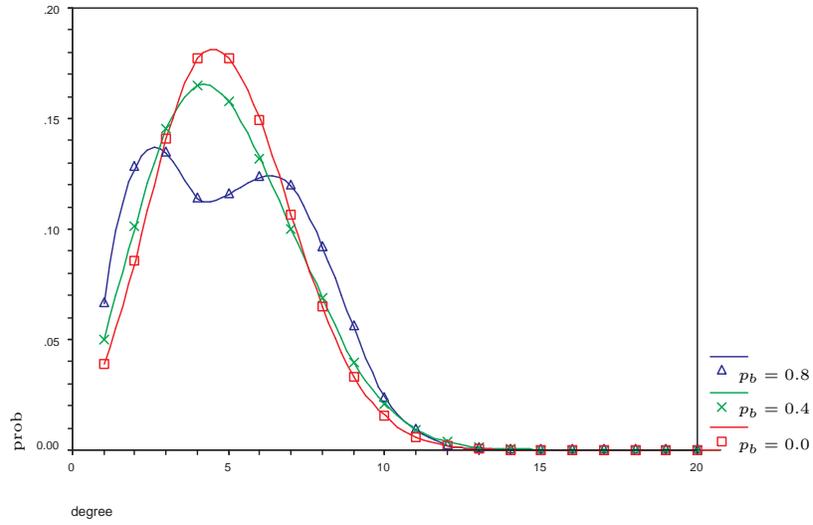} 
  \end{center}
\caption{\label{fig:degDistn} Average degree distribution for various values of $p_b$. 
}
\end{figure}

\begin{figure}
  \begin{center}
    \psfrag{Degree}{\tiny degree}
    \psfrag{Distance (scaled)}{\tiny scaled distance}
  \includegraphics[width=11cm]{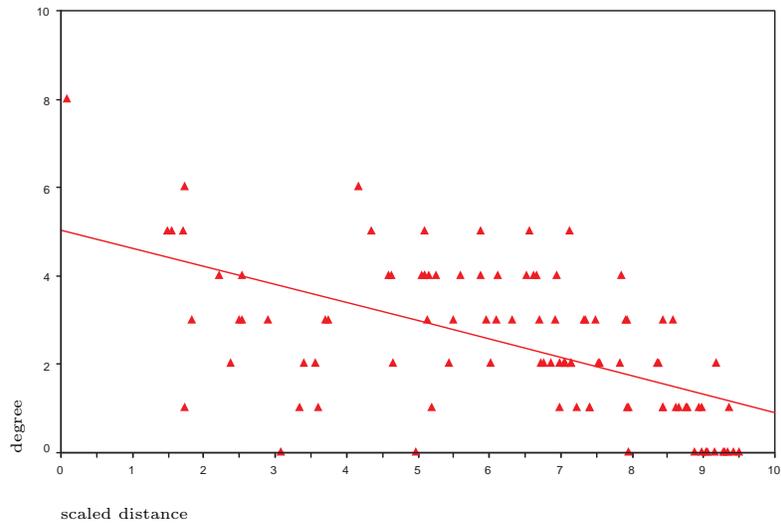}
  \end{center}
\caption{\label{fig:degcor} Scatterplot for the degree of nodes versus distance from the centre of the graph in one typical instance of the model ($N=100$, $p=0.05$, $p_b=0.8$).  The line is the least square fit line.  
}
\end{figure}

Fig.\ \ref{fig:degDistn} shows the average degree distribution of
graphs for various values of $p_b$.  They are all positively skewed
and as $p_b$ increases the corresponding degree distribution has an
increasingly fatter tail.  As $p_b$ becomes very large, the
distribution becomes a bimodal distribution.  This is due to the
boundary effect where nodes near the spatial boundary of the graph are
disadvantaged by having less possible neighbours within their
neighbourhood radius.  This is confirmed by the studying the
correlation between the spatial distance of nodes from the centre of a
typical graph and the degree of the nodes.  A scatterplot for a
typical instance can be found in Fig.\ \ref{fig:degcor} and in this
case it is found that there is a significant correlation between the
two ($p < 0.01$).

\section{An office communication network}
\begin{figure}
  \begin{center}
  \includegraphics[angle=-90,width=11cm]{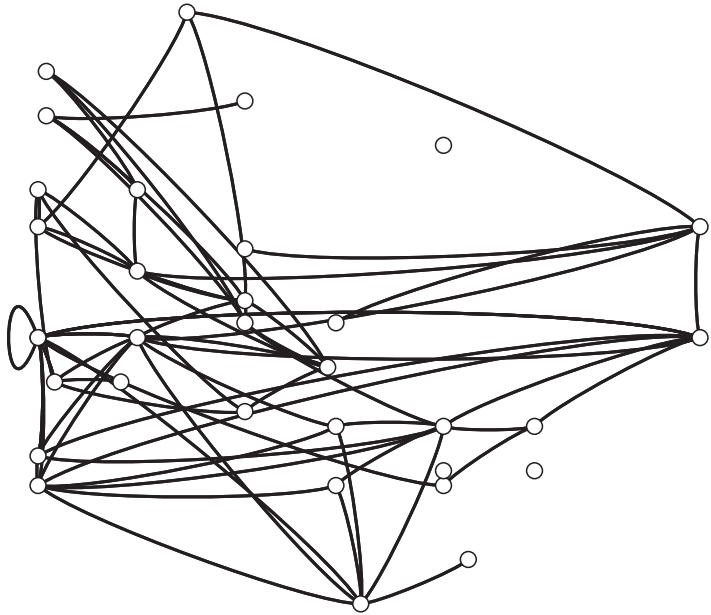}
  \end{center}
\caption{\label{fig:comnet-pajek} An office communication network of 33 individuals \cite{BerginRogers04}.  The location of the nodes reflects the cubicle or room locations.  Note that the ``self-loop'' in fact indicates a link between two individuals who share the same office space.
}
\end{figure}

We can use our current model to gain insights into the underlying
dynamics of a social network.  Our example here is a communication
network of 33 individuals observed over two days in a single-floor
office as part of a large organisation in Australia
\cite{BerginRogers04}.  We define a communication tie to exist between
two individuals if and only if each of the two parties has sought
information from each other more than three times over the two-day
period.  This definition is to avoid brief idiosyncratic encounters
that create much noise on top of the regular communication pattern.
The network is depicted in Fig.\
\ref{fig:comnet-pajek}.  The location of the nodes in the figure are
the actual locations of the individual's cubicle or room up to a
linear scaling.  In particular, the dimension of the office space is
scaled so that all relevant locations fits into a $1\times 1$ unit
square.

\begin{figure}
  \begin{center}
  \includegraphics[angle=-90,width=11cm]{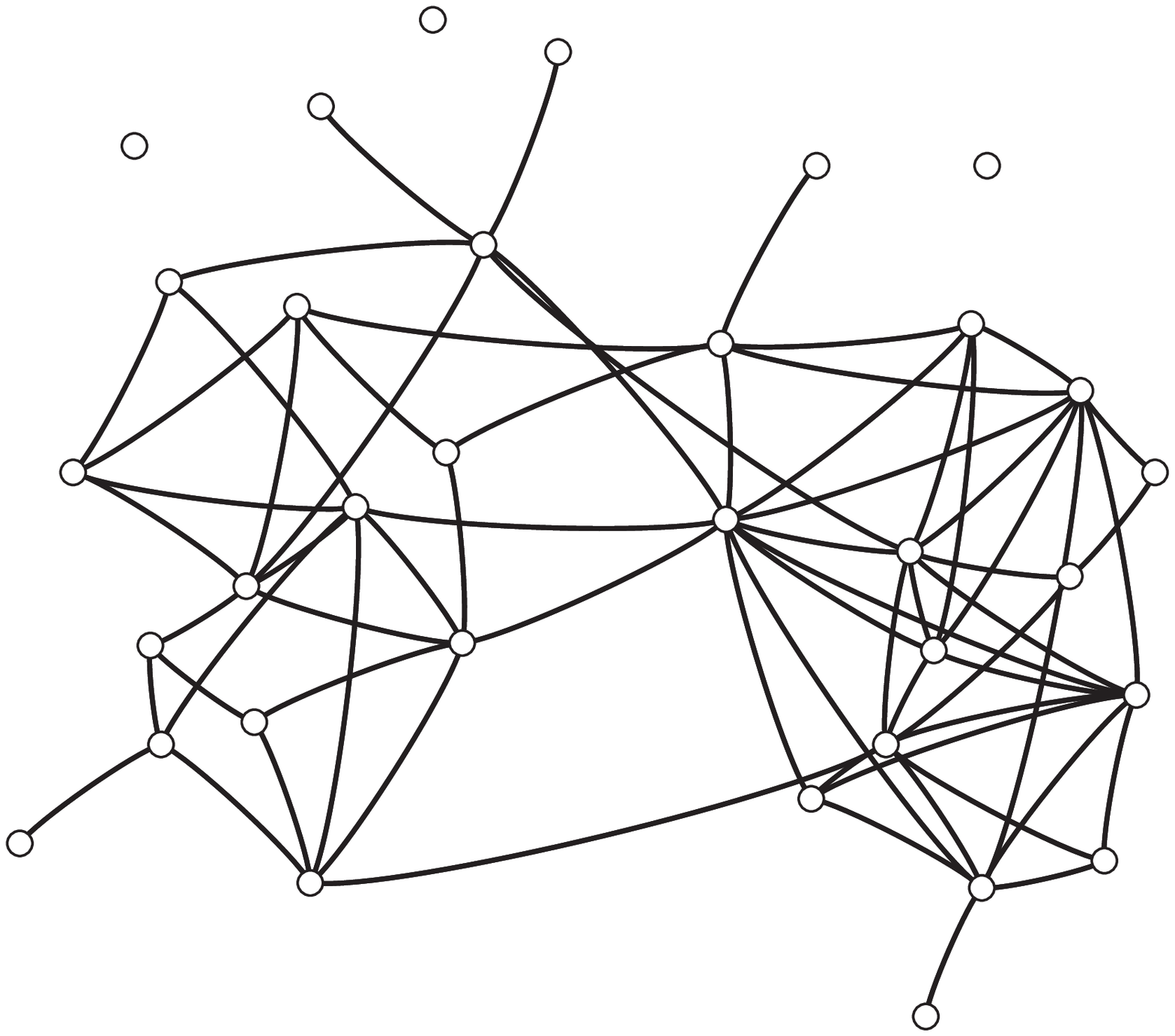} 
\end{center}
\caption{\label{fig:comnet-pajek-re} A spatial re-arrangement of the office communication network. 
}
\end{figure}

The network displays the generic features introduced in Section 1.2.
First of all, the density is low ($\rho = 0.131$).  The degree
distribution is (slightly) positively skewed (skewness statistics =
0.303).  The clustering coefficient is high (0.389) while the median
geodesic distance is very small (2).  Also, in a spatially rearranged
layout of the network in Fig.\ \ref{fig:comnet-pajek-re}, one can
easily identify the two distinct communities.

\begin{figure}
  \begin{center}
    \psfrag{p(d)}{$p(d)$}
    \psfrag{d}{$d$}
    \psfrag{empirical}{\tiny empirical}
    \psfrag{fitted}{\tiny fitted}
  \includegraphics[angle=-90,width=11cm]{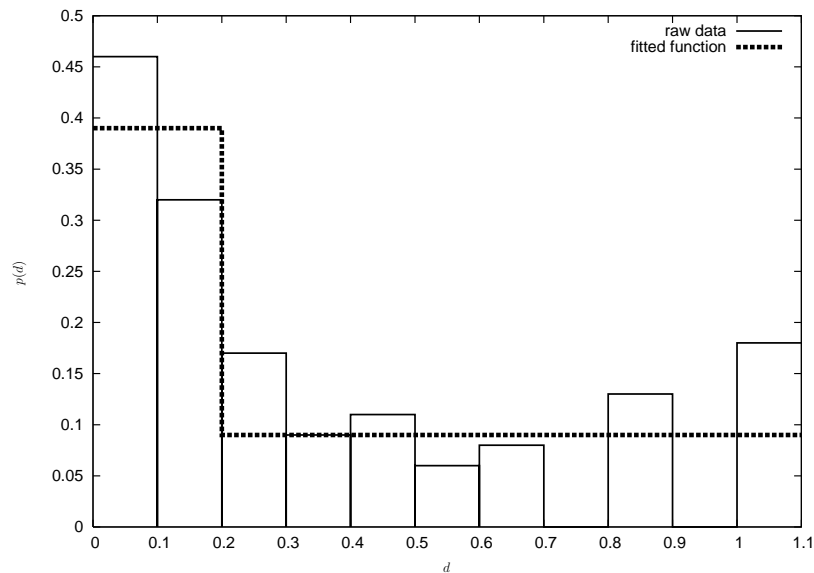}
  \end{center}
\caption{\label{fig:comnet-pd} A plot of edge probability $p(d)$ against distance $d$ in the communication network.  The solid bars are the empirical edge probability and the dashed line is the fitted step function with neighbourhood radius set at $H=0.2$.  Note that the distance has been scaled so that all node fits into a $1\times 1$ square.  All pairs of nodes are less than distance $1.1$ apart.
}
\end{figure}

Indeed, one can reasonably conjecture that spatial process is
important in this communication network.  To verify this, we use our
current model and estimate the model parameters, i.e.\ $p$, $H$, and
$p_b$ (see Eq.\ \ref{eq:modeldef}).  The empirical edge probability
$p(d)$ is plotted in Fig.\ \ref{fig:comnet-pd}.  It displays the
expected big drop at the small values of $d$.  Based on this, we can
fit $p(d)$ with a simple step function with some neighbourhood size
$H$ such that the sum of squared errors is minimal.  In this case, $H$
is found to be $0.2$.  As the overall density $p$ is $0.131$, the
proximity bias $p_b$ is estimated to be $0.259$, and the correction
term $\Delta$ is estimated to be $0.047$\footnote{Note that one can
alternatively derive the value of $\Delta$ using Eq.\
\ref{eq:delta}.}.  The large bias $p_b$ indicates that there is a
strong spatial component in the underlying social process.

\section{Conclusion}
In this study, we have performed Monte Carlo simulations on a class of
spatial random graph models.  It has been found that the properties of
these models differs significantly from the simple Erd\"{o}s-R\'{e}nyi
random graph model.  In particular, for a range of $p_b$ values, the
model displays simultaneously many general features of social networks
as discussed in Section 1 while the Erd\"{o}s-R\'{e}nyi random graph
model fails in many aspects.

We note that the properties of the model are similar to those of the
well-known Watts-Strogatz ``small world'' model
\cite{WattsStrogatz98} but that the current model has the advantage of
specifying an explicit probability distribution over the collection of
all graphs with a given number of nodes.  It is also important to note
that the clustering properties of the current model arise entirely
from the Poisson process describing node locations and hence from a
form of spatial baseline homophily.  It is clearly an empirical
question whether such models for social networks provide an adequate
descriptive account or whether it is necessary to also incorporate
inbreeding homophily effects (as in exponential random graph selection
models \cite{RobinsElliottPattison01}) or endogenous network processes
characteristic of the Watts-Strogatz model \cite{WattsStrogatz98} and
more general exponential random graph model specifications
\cite{BurdaJurkiewiczKrzywicki04,SnijdersPattisonRobinsHandcock05,WassermanPattison96,PattisonRobins02,HoffHandcockRaftery02}. This is a question that has received little attention in the network literature despite its fundamental importance to our understanding of
network evolution.

In order to identify models that provide a good match to empirical
data, it will be useful to construct a nested family of exponential
random graph models that can be used to evaluate the empirical
evidence for spatial and other forms of baseline homophily, inbreeding
homophily and endogenous clustering effects.  Indeed, reference to
empirical data immediately raises the possibility of at least two
alternative conceptualisations of a spatial model: the first, a
geographical space, in which geographical coordinates are associated
with each node; and the second, a more abstract ``social'' space, in
which spatial proximities reflect baseline homophily across a broad
range of individual attributes.  In each case, spatial locations may
be observed or unobserved.  For the current model and the case of
observed locations, it would be necessary to estimate the model
parameters $p$ and $H$ from the combination of location and network
data; in the case of unobserved locations, it would be desirable to
use a version of Hoff, Handcock and Raftery's
\cite{HoffHandcockRaftery02} more general exponential random graph
model specifications to estimate model parameters from network data
alone.  More generally, it would be desirable to construct within the
exponential random graph model family, models that also include
inbreeding homophily and endogenous clustering effects and associated
estimation methods.

The model described in this paper takes a useful first step towards
the construction of such a family of models.  The results presented
here suggest that the fit of models to empirical data will need
careful quantitative evaluation (e.g.\ as in
\cite{WassermanPattison96,SnijdersPattisonRobinsHandcock05}) because
of the likely capacity of many models within the family to exhibit in
broad terms the commonly observed features of empirical social
networks laid out in Section 1.2.

\section{Acknowledgments}
This study is financially supported by a Australian Research Council
Discovery Grant and the CSIRO--University of Melbourne Collaboration
Supportive Scheme.  We thank Sean Bergin and Paul Rogers for sharing
with us the communication network data.  We express our gratitudes to
Paul Walker, Stanley Wasserman and Tom Snijders for their helpful
discussions, and the anonymous referee for pointing out the relevant
literature and various other suggestions.  We would also like to thank
Peng Wang for programming the simulation program and the High
Performance Computer Facility at the University of Melbourne for
allowing us to use the IBM Alfred Cluster for simulations.

\newpage
\begin{thebibliography}{10}

\bibitem{AdamicAdar04} Adamic L., and Adar E., \textit{How to search a social network}, \textit{Preprint} (2004).

\bibitem{AlbertBarabasi02} Albert R., Barabasi A.-L., \textit{Statistical mechanics of complex networks}, Rev. Mod. Phys. \textbf{74}, 47-97 (2002). 

\bibitem{AmaralScalcBarthelemy00} Amaral L.A.N., Scala A., Barth\'{e}l\'{e}my M., and Stanley H.E., \textit{Classes of small-world networks}, Proc. Natl. Acad. Sci. USA \textbf{97}, 11149-11152 (2000).

\bibitem{AthanasiouYoshioka73} Athanasiou R. and Yoshioka G.A., \textit{The spatial characteristics of friendship formation}, Environment and behavior \textbf{5}, 43-66 (1973).

\bibitem{BarrettCampbell99} Barrett A.L. and Campbell K.E., \textit{Neighbor networks of black and white Americans}, in Networks in the global village: life in contemporary communities, Wellman B. (ed.), Westview Press (1999).

\bibitem{BerginRogers04} Bergin, S.\ and Rogers, P., \textit{Private communications}, Commonwealth of Australia (2004).

\bibitem{Bollobas85} Bollob\'{a}s B., \textit{Random Graphs}, Academic
Press, New York (1985).

\bibitem{BurdaJurkiewiczKrzywicki04} Burda Z., Jurkiewicz J., and Krzywicki A., \textit{Network transitivity and matrix models}, Phys.\ Rev.\ E \textbf{69}, 026106 (2004).

\bibitem{CaplowForman50} Caplow T. and Forman R., \textit{Neighbourhood interaction in a homogeneous community}, Am. Socio. Rev. \textbf{15}, 357-366 (1950).

\bibitem{CoxIsham80} Cox D.R. and Islam V., \textit{Point Processes}, Monographs on Applied Probability and Statistics, Chapman and Hall (1980).

\bibitem{Dunbar92} Dunbara R.I.M., \textit{Neocortex size as a constraint on group size in primates}, J. of Hum. Evolut. \textbf{20}, 469-493 (1992).

\bibitem{ErdosRenyi59} Erd\"{o}s P., and R\'{e}nyi A., \textit{On
random graphs.  I.}, Publicationes Mathematicae (Debrecen) \textbf{6},
290-297 (1959).

\bibitem{Feld81} Feld S.L., \textit{The focused organization of social ties}, Am. J. Soc. \textbf{86}, 1015-1035 (1981).

\bibitem{FestingerSchachterBack50} Festinger L., Schachter S., and Back K., \textit{Social processes in informal groups}, Standford Univ. Press (1950).

\bibitem{FrankStrauss86} Frank O. and Strauss D., \textit{Markov Graphs}, J. Am. Stat. Assoc. \textbf{81}, 832-842 (1986).

\bibitem{Granovetter73} Granovetter M.S., \textit{Strength of weak ties}, Am. J. Socio. \textbf{78}, 1360-1380 (1973).

\bibitem{HandcockJones03} Handcock, M., and Jones, J., \textit{An assessment of preferential attachment as a mechanism for human sexual network formation}, Proceedings of the Royal Society B \textbf{270}, 1123-1128 (2003).

\bibitem{HubermanAdamic04} Huberman, B.A., and Adamic L.A., \textit{Information Dynamics in the Networked World}, Lect.\ Notes Phys.\ \textbf{650}, Springer-Verlag Berlin Heidelberg, 371-398 (2004). 

\bibitem{HoffHandcockRaftery02} Hoff P.D., and Raftery A.E., and Handcock M.S., \textit{Latent Space Approaches to Social Network Analysis} J.\ Am.\ Stat.\ Assoc., \textbf{97}, 1090-1098 (2002).

\bibitem{JansonLuczakRucinski99} Janson S., Luczak T., and Rucinski A.
\textit{Random Graphs}, John Wiley, New York (1999).

\bibitem{Kingman93} Kingman J.F.C., \textit{Poisson Processes}, Oxford Studies in Probability 3, Clarendon Press, Oxford (1993).

\bibitem{Lazega01} Lazega E., \textit{The collegial phenomenon.  The social mechanisms of co-operation among peers in a corporate law partnership}, Oxford University Press, Oxford (2001). 

\bibitem{Logan78} Logan J.R., \textit{Growth, Politics, and the Stratification of places}, Am.\ J.\ Socio.\ \textbf{84}, 404-416 (1978).

\bibitem{McPhersonSmithLovinCook01} McPherson M., Smith-Lovin L., and Cook J.M., \textit{Birds of a Feather: Homophily in Social Networks}, Annu. Rev. Sociol \textbf{27}, 415-444 (2001).

\bibitem{Milgram67} Milgram S., \textit{The small world problem}, Psychology Today \textbf{2}, 60-67 (1967).

\bibitem{Morris97} Morris M., \textit{Sexual networks and HIV}, AIDS \textbf{11}, S209-S216 (1997).

\bibitem{MokWellmanBasu04} Mok D., Wellman B., and Basu R., \textit{Does distance matter for relationships?}, Presentation at SUNBELT International Social Network Conference, Portoroz, Slovenia (2004). 

\bibitem{Newman04} Newman M.E.J., \textit{Fast algorithm for detecting community structure in networks}, Phys. Rev. E \textbf{69}, 066133 (2004).

\bibitem{NewmanStrogatzWatts2001} Newman M.E.J., Strogatz S.H., and
Watts, D.J., \textit{Random graphs with arbitrary degree distributions
and their applications}, Phys.  Rev.  E \textbf{64} 026118 (2001).

\bibitem{Newman02} Newman M.E.J., \textit{The structure of scientific collaboration networks}, Proc. Natl. Acad. Sci. USA \textbf{98}, 404-409 (2001).

\bibitem{Newman03} Newman M.E.J., \textit{The structure and function of complex networks}, SIAM Review \textbf{45}, 167-256 (2003). 

\bibitem{Newman03a} Newman M.E.J., \textit{Mixing patterns in networks}, Phys. Rev. E \textbf{67} 026126 (2003). 

\bibitem{NewmanGirvan03} Newman M.E.J. and Girvan M., \textit{Mixing patterns and community structure in networks}, in Statistical Mechanics of Complex Networks, R. Pastor-Satorras, J. Rubi, and A. Diaz-Guilera (eds.), Springer, Berlin (2003). 

\bibitem{ParkNewman04} Park J. and Newman M.E.J., \textit{The statistical mechanics of networks}, Phys. Rev. E \textbf{70}, 066117 (2004).

\bibitem{PattisonRobins02} Pattison P.E., and Robins G. L., \textit{Neighbourhood-based models for social networks}, Sociological Methodology \textbf{32}, 301-337 (2002).   
\bibitem{Penrose91} Penrose M.D., \textit{On a continuum Percolation Model}, Adv. Appl. Prob. \textbf{23}, 536-556 (1991). 

\bibitem{Penrose03} Penrose M., \textit{Random Geometric Graphs}, Oxford University Press, Oxford (2003).

\bibitem{Rapoport53} Rapoport A., \textit{Spread of information through a population with socio-structural bias: I. assumption of transitivity}, Bull.\ Math.\ Biophys.\ \textbf{15}, 523-533 (1953).

\bibitem{Read54} Read K., \textit{Cultures of the central highlands, New Guinea}, Southwestern J. Anthro, 10, 1-43 (1954).

\bibitem{RobinsPattision05} Robins G.L., and Pattison P., \textit{Interdependencies and social processes: Generalized dependence structures}, In Carrington, Scott, and Wasserman (Eds) Models and Methods in Social Network Analysis.  Cambridge University Press (in press).

\bibitem{RobinsAlexander2004} Robins, G.L., and Alexander, M., \textit{Small worlds among interlocking directors: Network structure and distance in bipartite graphs}, Journal of Computational and Mathematical Organization Theory \textbf{10}, 69-94 (2004).

\bibitem{RobinsElliottPattison01} Robins G., Elliott P., Pattison, P., \textit{Network models for social selection processes}, Social Networks \textbf{23}, 1-30 (2001).

\bibitem{RobinsJohnston04} Robins G. and Johnston M., \textit{Joint social selection and social influence models for networks: The interplay of ties and attributes}, Presentation at SUNBELT International Social Network Conference, Portoroz, Slovenia (2004).

\bibitem{Snijders02} Snijders T.A.B., \textit{Markov Chain Monte Carlo Estimation of Exponential Random Graph Models}, J. Soc. Struc.\ \textbf{3}, No.\ 2 (electronic) (2002).

\bibitem{SnijdersPattisonRobinsHandcock05} Snijders T.A.B., Pattison P., Robins G., and Handcock, M., \textit{New Specification for exponential random graph models}, Socio. Meth., \textit{in press} (2005).

\bibitem{vanDuijnSnijdersZijlstra04} van Duijn M.A.J., Snijders T.A.B., and Zijlstra B.H., \textit{$p_2$: a random effects model with covariates for directed graphs}, Statistica Neerlandica \textbf{58} 234-254 (2004).

\bibitem{WattsStrogatz98} Watts D.J. and Strogatz S.H., \textit{Collective dynamics of `small-world' networks}, Nature \textbf{393}, 440-442 (1998).

\bibitem{Wellman96} Wellman B., \textit{Are personal communities local? A Dumptarian reconsideration}, Social Network \textbf{18}, 347-354 (1996).

\bibitem{WellmanCarringtonHall88} Wellman B., Carrington P., Hall A., \textit{Networks as personal communities}, in Social structures: A network approach, Wellman B. and Berkowitz S.D. (eds.), Cambridge University Press, Cambridge (1988).

\bibitem{WellmanWortley90} Wellman B., Wortley S., \textit{Different strokes from different folks: Community ties and social support}, American Journal of Sociology \textbf{96}, 558-588 (1990).

\bibitem{WassermanGalaskiewicz84} Wasserman S., Galaskiewicz J., \textit{Some generalizations of $p_1$ --- external constraints, interactions and non-binary relations}, Social Networks \textbf{6}, 177-192 (1984).

\bibitem{WassermanFaust94} Wasserman S.\ and Faust K., \textit{Social network analysis: Methods and applications}, Cambridge University Press, Cambridge (1994).

\bibitem{WassermanPattison96} Wasserman S. and Pattison P.,
\textit{Logit models and logistic regressions for social networks: I. An introduction to Markov Graphs and $p^*$}, Psychometrika \textbf{61}, 401-425 (1996).

\bibitem{Zipf49} Zipf G.K., \textit{Human behaviour and the principle of least effort}, Addison-Wesley (1949).

\end{thebibliography}
\end{document}